\documentclass[osajnl,twocolumn,showpacs,preprintnumbers,amsmath,amssymb]{revtex4}
\usepackage{graphicx}
\usepackage{dcolumn}
\usepackage{bm}

\def\stacksymbols #1#2#3#4{\def\theguybelow{#2}
        \def\verticalposition{\lower#3pt}
        \def\spacingwithinsymbol{\baselineskip0pt\lineskip#4pt}
        \mathrel{\mathpalette\intermediary#1}}
\def\intermediary#1#2{\verticalposition\vbox{\spacingwithinsymbol
        \everycr={}\tabskip0pt
        \halign{$\mathsurround0pt#1\hfil##\hfil$\crcr#2\crcr
                \theguybelow\crcr}}}

\def\gapproxeq{\stacksymbols{>}{\sim}{3}{.5}}

\begin{document}

\title{Controllable Raman-like nonlinearities from non-stationary cascaded quadratic processes}

\author{Fatih \"{O}. Ilday, Kale Beckwitt, Yi-Fan Chen, Hyungsik Lim, and Frank W. Wise}
\affiliation{Department of Applied and Engineering Physics, Cornell University, Ithaca, NY 14853.\\(607)255-9956 (phone), (607)255-7658 (fax), kb77@cornell.edu (e-mail)}

\begin{abstract}
We show that useful non-instantaneous nonlinear phase shifts can be obtained from cascaded quadratic processes in the presence of group velocity mismatch. The two-field nature of the process permits responses that can be effectively advanced or retarded in time with respect to one of the fields. There is an analogy to a generalized Raman-scattering effect, permitting both red and blue shifts of short pulses. We expect this capability to have many applications in short-pulse generation and propagation, such as the compensation of Raman-induced effects and high-quality pulse compression, which we discuss. 
\end{abstract}

\maketitle

\noindent

\section{Introduction}

In the last decade, there has been much interest in the nonlinear phase shifts produced by the cascaded interactions of two or three waves in quadratic ($\chi^{(2)}$) nonlinear media. Large nonlinear phase shifts of controllable sign can be generated, and numerous applications can be envisioned on the basis of such a capability \cite{cascade_review}. The prototypical quadratic process is second-harmonic generation (SHG). During the propagation of a fundamental-frequency (FF) field along with its second-harmonic (SH), the FF accumulates a nonlinear phase shift ($\Phi^{NL}$) if the process is not phase-matched. With long pulses (nanosecond-duration in practice) the FF and SH fields overlap temporally despite their different group velocities. In this so-called stationary limit, an effective Kerr nonlinearity is obtained (except at high intensity, when the fundamental field is depleted), and this can be a surrogate for the bound-electronic cubic ($\chi^{(3)}$) nonlinearity \cite{torner_cascade_background}. The cascade nonlinear phase shift can be thought of as arising from an effective nonlinear refractive index, {\em i.e.} the real part of an effective susceptibility. The residual SHG that occurs in the phase-mismatched process can similarly be considered the analog of two-photon absorption (the corresponding imaginary part of the effective susceptibility). 

The use of cascaded quadratic processes with ultrashort pulses is complicated substantially by the group-velocity mismatch (GVM) between the FF and SH fields \cite{Bakker, GVM_and_cascade}. GVM causes the fields to move apart in time, which reduces their coupling, and thus the magnitude of the cascade effects. In addition, the temporal profile of the nonlinear phase shift becomes distorted. Deviations of $\Phi^{NL}(t)$ from the pulse intensity profile hamper or preclude applications that involve soliton-like pulse shaping. The solution to this problem amounts to recovery of the stationary regime: For given value of the GVM, the phase mismatch is increased so that the cycles of conversion and back-conversion that generate the nonlinear phase shift occur before the pulses move apart from each other in time. Liu {\em et al.} showed that acceptable phase-shift quality can be obtained if at least 2 conversion cycles occur per characteristic GVM length $L_{\mathrm{GVM}} = 0.6 c \tau_{0}/(n_{g,1} - n_{g,2})$, which implies $\Delta k > 4 \pi/L_{\mathrm{GVM}}$ \cite{cascade_compression, wise_review}. Here, $c$ is the speed of light in vacuum, $\tau_{0}$ is the full-width at half-maximum (FWHM) of the pulse, and $n_{g,1}$ and $n_{g,2}$ are the group refractive indices for the FF and SH, respectively. In the limit of large phase mismatch an exact replica of a cubic nonlinearity is asymptotically obtained~\cite{torner_cascade_background}. The disadvantage of working with large phase mismatch is reduced magnitude of the nonlinear phase shift. GVM thus places a strong constraint on applications of cascade phase shifts. As one example, increasing GVM reduces the fraction of launched pulse energy that evolves into a soliton, eventually to zero \cite{acceptance_cascade}. To date, applications of cascade phase shifts with femtosecond pulses \cite{cascade_modelocking, cascade_compression, cascade_sfcomp, vortex_solitons} have all been demonstrated under stationary conditions. Approaching the stationary boundary, GVM coupled with self-phase modulation has been observed to asymmetrically broaden the pulse spectrum \cite{cascade_compression, pioger_spatialtrapping}

The nonlinear refraction experienced by an ultrashort pulse in a cubic nonlinear medium arises predominantly from bound-electronic and nuclear ({\em i.e.} Raman) contributions to the nonlinear response. Here we show that cascade phase shifts produced under non-stationary conditions mimic the Raman response, with some remarkable properties. Frequency shifts of controllable sign and magnitude can be impressed on short pulses. These effective Stokes and anti-Stokes Raman processes complete the analogy between cascade nonlinearities and true cubic nonlinearities, while maintaining the new degree of freedom provided by the quadratic interaction -- control of the process through the phase mismatch. An interesting feature of the non-stationary cascade process is that it provides a controllable non-instantaneous (and therefore nonlocal) nonlinearity. That the GVM alters the quadratic processes and produces deviations from a Kerr nonlinearity is well-known \cite{Bakker}. However, to date these effects have been perceived as distortions to be avoided. Just as the ability to control the sign and magnitude of an effective nonlinear index has enabled a new class of applications \cite{cascade_review, wise_review}, controllable Raman-like processes can be expected to create substantial new opportunities. We can think of cascade nonlinear processes with short pulses as dividing naturally into two classes, separated by the criterion for obtaining nonlinear phase shifts that mimic those of purely electronic origin. With this view, half of the possibilities have yet to be explored. Some examples will be discussed.

\section{Analytical Approach}

\begin{figure}[ht]
\centerline{\includegraphics[width=8.1cm]{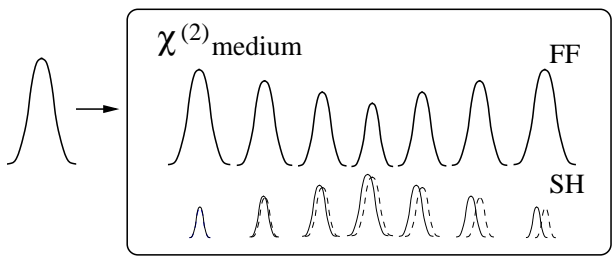}}
\caption{\label{cartoon} An illustration of the cascaded quadratic processes under phase-mismatched conditions. The FF is partially converted to the SH and then back-converted. Dashed (solid) lines are for the case of zero (nonzero) GVM.}
\end{figure}

The Kerr-like nonlinearity that arises from the $\chi^{(2)}:\chi^{(2)}$ process in the stationary limit can be understood qualitatively as follows. A small fraction of the FF is converted to the SH, which accumulates a phase difference before it is back-converted. The nonlinear phase shift impressed on the FF is delayed by one full cycle of conversion and back-conversion. However, as long as the SH is not displaced temporally from the FF, the phase shift on the FF will be proportional to its intensity profile (Fig.~\ref{cartoon}). With short pulses, the GVM becomes important if the fields separate by approximately the pulse duration before a cycle of conversion and back-conversion is complete. (We assume that the effect of the difference between the group-velocity dispersions (GVD) of the FF and the SH is negligible; in practice, it typically is much weaker than the inter-pulse GVM. This will be discussed quantitatively below.) Thus, after one cycle of conversion and back-conversion, the intensity profile of the SH field is retarded or advanced with respect to that of the FF, depending on the sign of the GVM. The corresponding delay of the nonlinear effect is slightly smaller or larger than one full conversion cycle. As a result, an effectively advanced or retarded phase shift is accumulated by the FF. The corresponding effect in the spectral domain is a frequency shift toward the blue or the red. Such frequency shifts have been predicted through numerical calculations~\cite{Bakker, acceptance_cascade}. 

The propagation of the FF and the SH are governed by coupled equations within the slowly varying envelope approximation (SVEA) \cite{coupled_eqns}. We neglect self- and cross-phase modulation due to $\chi^{(3)}$, consider only the temporal dimension, and assume conditions for type I second harmonic generation, but the results can easily be generalized.

\begin{eqnarray}
i \frac{\partial a_1}{\partial \xi} - \frac{\delta_1}{2} \frac{\partial^2 a_1}{\partial \tau^2} + a_1^* a_2 \exp(i \beta \xi) = 0, \label{prop_eqn_ff}\\
\nonumber\\
i \frac{\partial a_2}{\partial \xi} - \frac{\delta_2}{2} \frac{\partial^2 a_2}{\partial \tau^2} - i \frac{\partial a_2}{\partial \tau} + a_1^2 \exp(-i \beta \xi) = 0. \label{prop_eqn_sh} \\ \nonumber
\end{eqnarray}

Here $a_1$ and $a_2$ are the normalized FF and SH field amplitudes. Time is normalized to the initial pulse duration, $\tau = t/\tau_0$, and the scaled propagation coordinate is $\xi = z/L_{\mathrm{GVM}}$. Here, $\delta_j = L_{\mathrm{GVM}} / L_{DS,j}$, where $L_{DS,j}=0.322 {\tau_0}^2/\mathrm{GVD}(\omega_j)$ are the dispersion lengths with $j = 1, 2$. The parameter $\beta = \Delta k L_{\mathrm{GVM}}$ where $ {\Delta}k = k_{2\omega}-2k_{\omega}$ is the normalized FF-SH wave-vector mismatch.

Consider the simple but common case when only FF light is incident on the quadratic medium. In the limit of large phase-mismatch, conversion and back-conversion occur rapidly and most of the energy resides in the FF at all times. A relation between the FF and the SH amplitudes can be derived as an expansion in powers of $\beta$~\cite{torner_cascade_background}. By eliminating $a_2$ in Eq.~\ref{prop_eqn_sh} and keeping terms up to order $(1/\beta^3)$, an equation of motion for the FF field can be derived:

\begin{eqnarray}
i \frac{\partial a_1}{\partial \xi} - \frac{\delta_1}{2} \frac{\partial^2 a_1}{\partial \tau^2} - \frac{1}{\beta} |a_1|^2 a_1 - 2 i \frac{1}{\beta^2} |a_1|^2 \frac{\partial a_1}{\partial \tau}\; \nonumber \\
\nonumber \\
+ \frac{1}{\beta^2} (\delta_1 - \delta_2 ) |a_1|^2 \frac{\partial^2 a_1}{\partial \tau^2} - \frac{\delta_2}{\beta^2} a^{*}_{1} (\frac{\partial a_1}{\partial \tau})^2+ O(\frac{1}{|{\beta}|^3}) = 0.\; \label{NLSE-1} \\ \nonumber
\end{eqnarray}

The first three terms constitute a nonlinear Schr\"{o}dinger equation (NLSE), which is generalized by the fourth and the fifth terms that describe the effects of the GVM, and the mismatch of the GVD of the FF and SH, respectively. The final term is negligible (at order $1/\beta^2$) since it is proportional to the square of the first derivative of the field envelope (a small quantity in the SVEA). In addition it is even in time for well behaved fields ($a_1(t)$), and hence cannot contribute to the frequency shifting process. We neglect the GVD mismatch since its effect is much smaller than inter-pulse GVM for typical nonlinear media under the assumed conditions of small conversion to the SH. In that case, Eq.~\ref{NLSE-1} reduces to:

\begin{eqnarray}
 i \frac{\partial a_1}{\partial \xi} - \frac{\delta_1}{2} \frac{\partial^2 a_1}{\partial \tau^2} - \frac{1}{\beta} |a_1|^2 a_1 - 2 i \frac{1}{\beta^2} |a_1|^2 \frac{\partial a_1}{\partial \tau} = 0. \label{NLSE-2} \\ \nonumber
\end{eqnarray}

We notice that Eq.~\ref{NLSE-2} resembles the Chen-Liu-Lee equation (CLLE), which is integrable~\cite{CLLE}. The difference is the presence of a cubic nonlinear term. It can be shown that Eq.~\ref{NLSE-2} reduces to the CLLE:

\begin{eqnarray}
i \frac{\partial q(Z,T)}{\partial Z} - \frac{\partial^2 q(Z,T)}{\partial T^2} + i |q(Z,T)|^2 \frac{\partial q(Z,T)}{\partial T} = 0, \\ \nonumber
\end{eqnarray}

with the substitution, 

\begin{eqnarray}
a_1(\xi,\tau) = c_{0} \exp[i(c_1 \xi + c_2 t)] q(\xi, \tau), \label{substitution} \\ \nonumber
\end{eqnarray}

where $|c_{0}|^2 = \beta^2 \sqrt{\delta_{1}/8}$, $c_{1} = \delta_{1} \beta^{2}/8$, and $c_{2} = \beta/2$, followed by the coordinate transformation 

\begin{eqnarray}
T = - \sqrt{2/\delta_{1}}\; \tau - \beta \sqrt{\delta_{1} /2}\; \xi, \nonumber \\
 \label{transformation} \\
Z = \xi. \nonumber  \\ \nonumber
\end{eqnarray}

Hence, Eq.~\ref{NLSE-2} is integrable as well. We note that the CLLE is further related to the well-known, integrable derivative nonlinear Schr\"{o}dinger equation, via a Gauge-invariant transformation~\cite{CLLE2DNLSE}. 

To further understand the effects of the lowest order correction from GVM in Eq.~\ref{NLSE-2}, we can compare it to the equation governing the propagation of a single field envelope $a_1$ under similar approximations to Eqs.~\ref{prop_eqn_ff}-\ref{prop_eqn_sh}, but for a Kerr-nonlinear material with finite Raman-response time $T_R$: \cite{Agrawal}

\begin{eqnarray}
\underbrace{i \frac{\partial a_1}{\partial \xi} - \frac{\delta_1}{2} \frac{\partial^2 a_1}{\partial \tau^2} + \gamma \big[ |a_1|^2 a_1}_{\mathrm{NLSE}} - \underbrace{T_R  a_1 \frac{\partial |a_1|^2}{\partial \tau}}_{\mathrm{Raman}} \big] = 0. \label{NLSE-Raman} \\ \nonumber
\end{eqnarray}

Here $\gamma = n_2 \omega_0 / (c \pi w^2)$ for a Gaussian beam with of frequency $\omega_0$ and waist $w$. Comparing Eq.~\ref{NLSE-Raman} to Eq.~\ref{NLSE-2}, we see that they are similar but with the Raman term of Eq.~\ref{NLSE-Raman} replaced by $T_R^{\mathrm{eff}} |a_1|^2 \partial a_1/\partial \tau$ where $T_R^{\mathrm{eff}} \equiv -2i/\beta$. While the correspondence is not exact since the functional dependence is different ($\sim a_1 \partial |a_1|^2/\partial \tau$ for Raman-scattering vs. $\sim |a_1|^2 \partial a_1/\partial \tau$ for the cascaded process), some qualitative understanding can be gained from considering the effective cascaded response with $T_R^{\mathrm{eff}} \sim i/\beta$: first, the cascaded correction is imaginary and hence does not contribute directly to the phase (unlike the Raman response). Rather it alters the field envelope. The envelope change subsequently couples to the phase profile through the remaining terms of Eq.~\ref{NLSE-2}, so the frequency shift occurs through a higher order process. Second, $T_R^{\mathrm{eff}}$ saturates with $1/\beta \sim 1/{\Delta}k(\lambda)$, unlike the Raman response which does not depend strongly on wavelength. This saturation will be explored in greater detail in Section \ref{num_analysis}. Note that the effective response for the cascaded process can be $\gapproxeq$ two orders of magnitude greater than that of Raman-scattering, so that significant frequency shifting is possible in cm of quadratic material (vs. meters of fiber with Raman).

For a qualitative understanding of the effect of the GVM term, we decompose the field $a_1(\xi, \tau)$ in Eq.~\ref{NLSE-2} into its amplitude and phase with the substitution $a_1(\xi, \tau) = u(\xi, \tau) \exp(i \phi(\xi, \tau))$, where $u(\xi, \tau)$ and $\phi(\xi, \tau)$ are real functions. The evolution of the amplitude and the phase is then given by

\begin{eqnarray}
\frac{\partial u}{\partial \xi} = \delta_1 \frac{\partial u}{\partial \tau} \frac{\partial \phi}{\partial \tau} + \frac{\delta_1}{2} u \frac{\partial^2 \phi}{\partial \tau^2} + \frac{2}{\beta^2} u^2 \frac{\partial u}{\partial \tau},\; \nonumber \\
\; \\
u \frac{\partial \phi}{\partial \xi} = - \frac{\delta_1}{2} \frac{\partial^2 u}{\partial \tau^2} + \frac{\delta_1}{2} (\frac{\partial \phi}{\partial \tau} )^2 u - \frac{1}{\beta} u^3 + \frac{2}{\beta^2} u^3 \frac{\partial \phi}{\partial \tau}.\; \nonumber \\ \nonumber
\end{eqnarray}

For the sake of simplicity, we concentrate on the nonlinear terms and ignore the dispersion, which amounts to neglecting the terms with higher-order time derivatives. With this simplification, we obtain

\begin{eqnarray}
\frac{\partial u}{\partial \xi} = \frac{2}{\beta^2} u^2 \frac{\partial u}{\partial \tau},\; \label{amplitude} \\
\nonumber \\
u \frac{\partial \phi}{\partial \xi} = - \frac{1}{\beta} u^3 + \frac{2}{\beta^2} u^3 \frac{\partial \phi}{\partial \tau}.\; \label{phase}  \\ \nonumber
\end{eqnarray}

If we assume $\beta$ is large, $u(\xi, \tau)$ and $\phi(\xi, \tau)$ can be calculated by expanding in powers of $\frac{1}{\beta}$, similar to the procedure used to obtain Eq.~\ref{NLSE-2}. Keeping terms up to order $\frac{1}{\beta^2}$, Eq.~\ref{amplitude} shows that $u(\xi, \tau) = u(\tau) + O(\frac{1}{\beta^2})$, that the field amplitude is approximately unchanged. Thus, integration of Eq.~\ref{phase} yields

\begin{eqnarray}
\phi(\xi, \tau) = -\frac{1}{\beta} u^2(\tau) \xi + O(\frac{1}{\beta^2}). \\ \nonumber
\end{eqnarray}

Substitution of this relation back into Eq.~\ref{phase} gives

\begin{eqnarray}
\frac{\partial \phi}{\partial \xi} \approx - \frac{1}{\beta} u^2 - \frac{4}{\beta^3} u^3 \frac{\partial u}{\partial \tau}.\; \label{phase-alone}  \\ \nonumber
\end{eqnarray}

The first term on the right represents the Kerr-like nonlinear phase shift and the second term corresponds to the non-instantaneous nonlinear response due to large GVM.

The relation in Eq.~\ref{phase-alone} provides a valid description of the phase evolution only in its early stages, before the field amplitude is modified significantly, and in the absence of dispersion. Within these approximations, the effect of the GVM on the nonlinear phase shift can be illustrated for a given pulse shape: $a_1(0,\tau) = {\rm sech}(\tau)$. Integration of Eq.~\ref{phase-alone} yields 

\begin{eqnarray}
\phi(\xi, \tau) \approx - \frac{1}{\beta} {\rm sech}^2(\tau) (1 + \xi \frac{2}{\beta^2} {\rm sech}(\tau) \tanh(\tau)) \xi + \phi_{0},\; \\ \nonumber
\end{eqnarray}

where $\phi_{0}$ is an integration constant. The temporal asymmetry of the GVM contribution shifts the peak of the nonlinear phase shift.

Likewise, Fourier transforming to the frequency domain, the contributions of the Kerr-like and the GVM terms in Eq.~\ref{NLSE-2} can be calculated for the pulse shape to be 

\begin{eqnarray}
\frac{1}{\beta} |a_1|^2 a_1 + 2 i \frac{1}{\beta^2} |a_1|^2 \frac{\partial a_1}{\partial \tau} = \; \nonumber \\
\nonumber \\
\frac{\sqrt{\pi/2}}{\beta} (1 + \omega^2) (3 + 2 \omega/\beta) {\rm sech}(\pi \omega/2),\; \\ \nonumber
\end{eqnarray}

which has a bipolar shape. Positive frequency components are attenuated and negative frequencies are amplified, or vice versa, depending on the sign of the phase-mismatch to GVM ratio. This result is expected to hold in general for any smooth, single-peaked pulse shape for which $a_1(0,\tau) \rightarrow 0$ for $|\tau| \rightarrow \infty$. Such a frequency shift is expected from the GVM term, which has an odd-order time derivative.

\section{Numerical Analysis \label{num_analysis}}

Although the approximate one-field equation (Eq.~\ref{NLSE-2}) is useful for a qualitative understanding, it is necessary to consider the coupled equations (Eqs.~\ref{prop_eqn_ff}-\ref{prop_eqn_sh}) for a quantitative description. To this end, we numerically solve a version of Eqs.~\ref{prop_eqn_ff}-\ref{prop_eqn_sh} that has been generalized to include self- and cross-phase modulation terms due to the cubic nonlinearity. We use a different field normalization here to facilitate comparison to experimental parameters.

\begin{eqnarray}
\lefteqn{i \frac{\partial A_1}{\partial z} - \frac{Z_I}{2 L_{DS,1}}\frac{\partial^2 A_1}{\partial \tau^2}} \nonumber \\
\nonumber \\
&&{} + A_1^* A_2 \exp(i \Delta k (Z_I z)) \nonumber \\
\nonumber \\
&&{} + \frac{Z_I}{L_{NL,1}} (|A_1|^2 + 2 |A_2|^2)A_1 = 0, \label{prop_eqn_full_ff} \\
\nonumber \\
\lefteqn{i \frac{\partial A_2}{\partial z} - \frac{Z_I}{2 L_{DS,2}}\frac{\partial^2 A_2}{\partial \tau^2} - i \frac{Z_I}{L_{\mathrm{GVM}}}\frac{\partial A_2}{\partial \tau} } \nonumber \\
\nonumber \\
&&{} + \frac{n(\omega_1)}{n(\omega_2)} A_1^2 \exp(-i \Delta k (Z_I z)) \nonumber \\
\nonumber \\
&&{} + \frac{n(\omega_1)}{n(\omega_2)} \frac{Z_I}{L_{NL,2}} (2 |A_1|^2 + |A_2|^2)A_2 = 0. \label{prop_eqn_full_sh} \\ \nonumber
\end{eqnarray}

Here the FF and SH envelopes ($A_1$ and $A_2$, respectively) are in units of the initial peak FF field $A_0$ (related to the initial peak FF intensity by $I_0=\sqrt{\varepsilon/\mu}|A_0|^2/2$) and $n_2$ is the Kerr nonlinear index. The cubic nonlinear length characterizing the pulse propagation is $L_{NL,j} = c/\omega_j n_2 I_0$ (the length over which the accumulated nonlinear phase shift is 1) for frequency $\omega_j$ with $j = 1,2$. The propagation coordinate $z$ is normalized to the quadratic nonlinear length $Z_I =n\lambda_1/{2\pi d_{\mathrm{eff}} A_0}$ which characterizes the strength of the nonlinear coupling. $d_\mathrm{eff}$ is the effective quadratic nonlinear coefficient. Time $\tau$, $L_{DS,j}$, and $L_{\mathrm{GVM}}$ are defined as for Eqs.~\ref{prop_eqn_ff}-\ref{prop_eqn_sh}.

\begin{figure}[ht]
\centerline{\includegraphics[width=8.1cm]{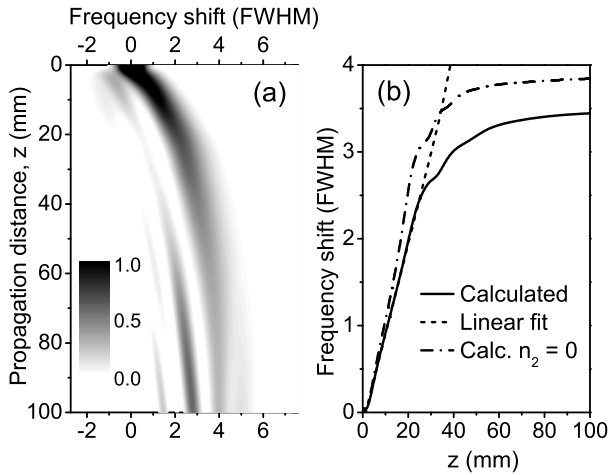}}
\caption{\label{freq_shift_sat} (a) Evolution of the spectrum along the propagation direction. Shift is in units of the initial spectral FWHM, $\sim$ 3.7 THz. The scale bar shows spectral intensity in arb. units. (b) Weighted average frequency shift as a function of propagation distance. Dashed line indicates fit to region of linear shift. Dash-dots show similar results in the absence of $\chi^{(3)}$ ($n_2 = 0$).}
\end{figure}

Eqs.~\ref{prop_eqn_full_ff}-\ref{prop_eqn_full_sh} are solved using a symmetric split-step beam propagation method~\cite{cascade_sts}. The simulations assume typical conditions for femtosecond pulses in quadratic nonlinear crystals. As an example, we calculate the propagation of 120-fs pulses with a peak intensity of $50\;{\rm GW/cm^{2}}$ in a 10 cm-long quadratic medium. The launched pulse shape is chosen to be Gaussian and the center wavelength is 790 nm. The quadratic medium used is barium metaborate (${\rm Ba_{2}BO_{4}}$ or BBO), for which the material parameters are: n = 1.63, $d_{\mathrm{eff}}$ = 1.82 pm/V, $n_2 = 3.2 \times 10^{-16} {\;}\mathrm{cm^2/W}$, GVM = -186 fs/mm, the FF (SH) GVD = 70 (190) $\mathrm{fs^2/mm}$, and the FF (SH) third-order dispersion is 50 (81) $\mathrm{fs^3/mm}$. Note that the true cubic nonlinearity ($n_2$) is included in the calculations. Two-photon absorption is neglected since it is small for BBO at 790 nm. The phase-mismatch is set to be ${\Delta}k = 5 \pi/\mathrm{mm}$, corresponding to a self-defocusing nonlinearity and a magnitude that is about half the minimum value to obtain a Kerr-like phase shift. This particular set of conditions, except for the crystal length, is chosen to correspond to experiments that are described later in the text. As expected, the spectrum of the pulse shifts to higher frequencies as it propagates through the quadratic medium. The evolution of the spectrum is shown in Fig.~\ref{freq_shift_sat}(a). Initially, the frequency-shift increases linearly with propagation distance, but eventually the process saturates (Fig.~\ref{freq_shift_sat}(b)). The spectrum of the SH field shifts opposite from the FF (\emph{i.e.} to lower frequencies), prior to saturation. This saturation is expected, since the effective response of the cascaded process decreases (or is distorted) with increases frequency shift. For this choice of pulse parameters and phase-mismatch, the saturation begins beyond 3 cm, which is close to the maximum length of available BBO crystals. The cubic electronic nonlinearity of the quadratic material is included here for complete correspondence with experimental parameters, however the close agreement in Fig.~\ref{freq_shift_sat}(b) between the saturation trend with and without $n_2$ indicates the dominance of the quadratic process in the frequency shifting dynamics. In Fig~\ref{freq_shift_sat}(b), the presence of cubic nonlinearity slightly reduces the resulting frequency shift, as expected due to its self-focusing phase. One might expect the material's Raman response to be relevant to the frequency shifting dynamics studied here, however the Raman response of BBO with $\sim$100-fs pulses is small compared to the cascaded response. In comparison to Raman, the cubic electronic nonlinearity included above in Fig.~\ref{freq_shift_sat}(b) is a larger effect, even though it alters the frequency shifting process indirectly through the nonlinear phase.

\begin{figure}[ht]
\centerline{\includegraphics[width=8.1cm]{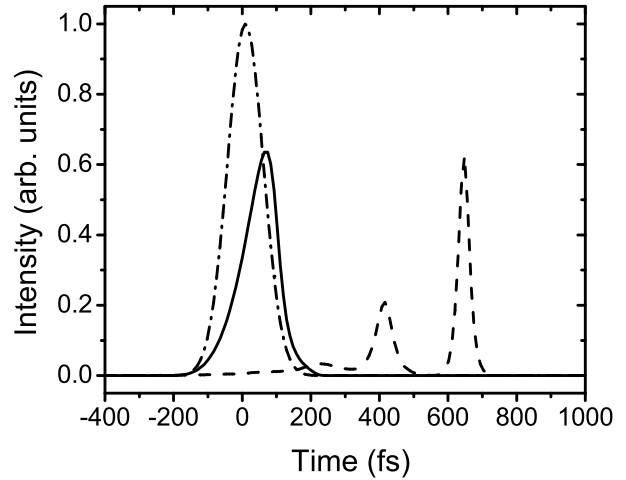}}
\caption{\label{temporal_profile} Intensity profiles of the FF at $z = 0\; {\rm mm}$ (dash-dotted line), $z = 12\; {\rm mm}$ (solid line), $z = 100\; {\rm mm}$ (dashed-line). For the launched pulse $L_{DS,1} = 74\;{\rm mm}$.}
\end{figure}

The pulse propagation is dominated by an interplay of GVD and the effective nonlinearity from the cascaded process in the form of soliton-like dynamics. In the time domain, the pulse undergoes compression since the energy is more than the amount needed to balance the dispersive effects. The intensity profiles before the onset of saturation and at the exit face of the crystal are plotted in Fig.~\ref{temporal_profile} along with those of the launched pulse. The FF undergoes a steady compression accompanied by energy loss to the SH: At $z = 12\; {\rm mm}$, its FWHM is 110 fs with $64 \%$ of the pulse energy remaining in the FF. With propagation, the pulse shape becomes slightly asymmetrical. The asymmetrical structure develops into a secondary pulse in the final stages of the propagation, which corresponds to the secondary structure of the spectrum (Fig.~\ref{freq_shift_sat}(a)). At $z = 100\; {\rm mm}$, the FWHM of the main peak is reduced to 40 fs while $\sim 36 \%$ of the launched energy is retained in the FF. The temporal profiles are displaced since the pulse experiences different group velocities as its central frequency changes.

\begin{figure}[ht]
\centerline{\includegraphics[width=8.1cm]{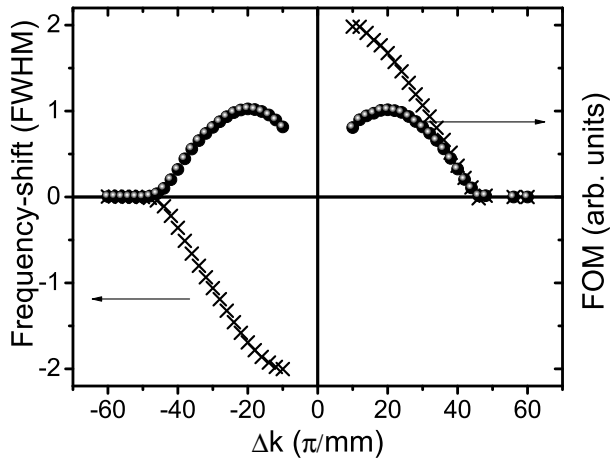}}
\caption{\label{FOM_fig} Frequency shift (crosses) and FOM (circles) as a function of phase-mismatch. Similarly to Fig.~\ref{freq_shift_sat}, frequency shift is measured in units of the initial FWHM (here $\sim$ 4.4 THz). Note that GVD is chosen to be normal (anomalous) for $\Delta k > 0$ ($\Delta k < 0$) to support soliton-like pulses.}
\end{figure}

A similar picture emerges for the total frequency shift for fixed propagation distance and varying phase-mismatch. The magnitude of the nonlinear phase, and hence the frequency shift, is proportional to $1/\Delta k$ before it saturates. Loss to SH conversion increases with decreasing $|\Delta k|$, so there exists a trade-off between the magnitude of the frequency shift and loss. We define a figure-of-merit (FOM) for the shifting process as the ratio of frequency-shift to energy content in the SH field, which attains a maximum for phase-mismatch values slightly below that of the minimum for obtaining Kerr-like nonlinear phase shifts (Fig.~\ref{FOM_fig}). This is demonstrated in Fig.~\ref{FOM_fig}, which shows simulations of 100-fs, 200-pJ pulses with center wavelength 1550 nm (and peak intensity $\sim 5 {\;}\mathrm{GW/cm^2}$). The material parameters used correspond to the quadratic material periodically-poled lithium niobate (PPLN): n = 2.14, $d_{\mathrm{eff}}$ = 16.5 pm/V, $n_2 = 3.2 \times 10^{-15} {\;}\mathrm{cm^2/W}$, GVM = -370 fs/mm, and the FF (SH) GVD = 100 (400) $\mathrm{fs^2/mm}$. Under these conditions, the stationary boundary for Kerr-like phase shifts corresponds to $|{\Delta}k| \gapproxeq 25\pi/\mathrm{mm}$. Notice that much larger frequency shifts can be generated closer to phase-matching, but with larger SH-conversion.

\begin{figure}[ht]
\centerline{\includegraphics[width=8.1cm]{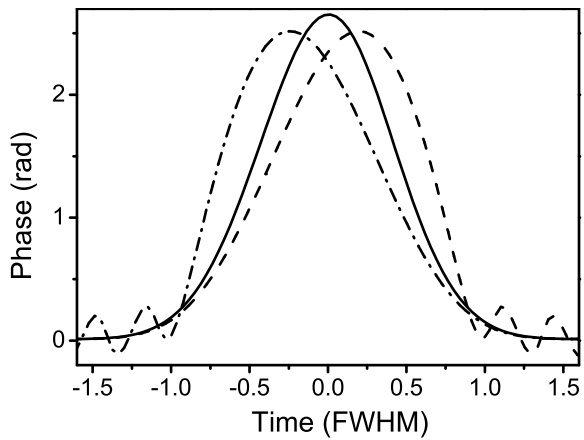}}
\caption{\label{non-inst} The phase impressed on the FF for zero (solid line), positive (dashed line), and negative (dash-dotted line) GVM.}
\end{figure}

The non-instantaneous nature of the cascaded quadratic process with significant walk-off between the FF and the SH is demonstrated by the nonlinear phase shift imposed on the FF. Simulations confirm the aforementioned expectations: Effectively retarded or advanced phase shifts are imposed on the FF depending on the sign of the GVM (Fig.~\ref{non-inst}).

\section{Experimental Observation of the Frequency-shift}

Experiments were performed with 120-fs, 0.6-mJ pulses, centered at 790 nm, generated by a Ti:sapphire regenerative amplifier. The launched pulse shape was approximately Gaussian with a clean spatial profile and the peak intensity is estimated to be $50\;{\rm GW/cm^{2}}$. A 17 mm-long piece of BBO served as the quadratic medium. The GVM length is $L_{\mathrm{GVM}}$ = 0.38 mm, for which the criterion for a Kerr-like phase shift implies ${\Delta}k > 10.4 \pi{\rm /mm}$. Both blue and red shifts are experimentally available via positive and negative phase-mismatch, respectively. However, red-shifts occur with self-focusing nonlinearity, which limits the peak intensity available without continuum generation and crystal damage. Consequently we focus here on blue-shifts. 

\begin{figure}[ht]
\centerline{\includegraphics[width=8.1cm]{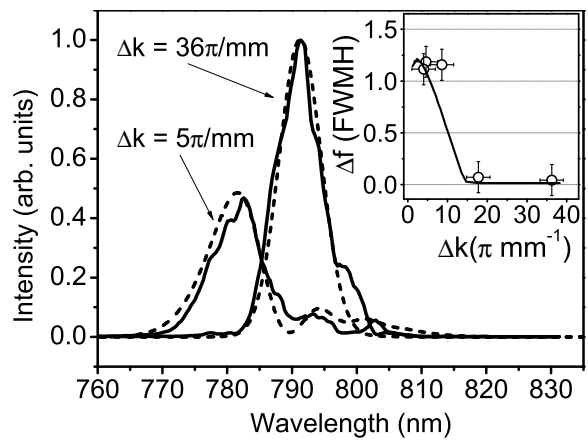}}
\caption{\label{shift_fig} Experimental (solid lines) and simulated (dashed lines) spectra for phase-mismatches of $5 \pi/\mathrm{mm}$ and $36 \pi/\mathrm{mm}$. The latter serves as control. Inset: Experimental (symbols) and calculated (solid line) frequency shift for different values of phase-mismatch. As in Fig.~\ref{freq_shift_sat}, frequency shift is measured in units of the initial FWHM, $\sim$ 3.7 THz.}
\end{figure}

Increasing frequency-shift of the FF was observed with decreasing phase-mismatch (Fig.~\ref{shift_fig}). The inset of Fig.~\ref{shift_fig} shows the spectral shift versus phase-mismatch for self-defocussing phase shifts, and the main figure shows example spectra. The data presented for ${\Delta}k = 36 \pi/\mathrm{mm}$ serve as a control experiment: At such a large phase mismatch the cascade nonlinear phase and the Kerr nonlinearity are negligible and the spectrum is indistinguishable from the spectrum of the launched pulses (not shown). The temporal profile of the pulse did not change significantly in these experiments. The experimental results are compared to the results of numerical simulations which contained no free parameters and were based on experimental conditions. In Fig.~\ref{shift_fig}, the calculated spectrum of the unshifted pulse ($36 \pi/\mathrm{mm}$) is normalized so that it contains the same power as the measured unshifted spectrum and all other traces have the same relative scaling, so that the units of all the given spectra are the same. With this in mind, there is excellent agreement between the measured spectra and the simulations. In particular, the shift increases greatly from ${\Delta}k = 19 \pi/\mathrm{mm}$ to ${\Delta}k = 5 \pi/\mathrm{mm}$. The latter is the phase-mismatch for which the ratio of the spectral-shift to SH conversion should peak. The dependence of the frequency-shift on the phase-mismatch, as summarized by the inset of Fig.~\ref{shift_fig}, agrees qualitatively with the results of Fig.~\ref{FOM_fig}; however, the phase-mismatch corresponding to maximum spectral-shift to SH conversion is different than in Fig.~\ref{FOM_fig} as a consequence of different physical parameters in the experiment.

\section{Applications}

We have shown that cascaded quadratic processes under phase-mismatched conditions and in the presence of significant GVM (typical conditions with femtosecond-duration pulses) result in an effectively non-instantaneous cubic nonlinearity. The response time is controllable by appropriate choice of the phase-mismatch. In addition to providing an intuitive picture for the effect of phase-mismatch on the propagation of femtosecond pulses, this nonlinear process offers some unique features. The nonlocal nature of the Raman-like cascade nonlinearity is interesting in its own right. Non-locality of the cubic nonlinearity has been shown to arrest self-focusing collapse and to stabilize solitons, for example \cite{nonlocal_no_collapse, nonlocal_NL_review}. The nonlocal nature of the cascade process under non-stationary conditions can be controlled or tailored to specific situations via the phase-mismatch.

Many applications of a controllable effective Raman process can be envisioned. Perhaps the most obvious one is the cancellation of the Raman shift that a short pulse accumulates as it propagates in optical fiber. For example, in telecommunication systems with bit rates above $\sim 20$ Gbit/s, the pulse duration is short enough that timing jitter is dominated by jitter arising from Raman-induced frequency shifts \cite{Raman_communication_limit}. 

\begin{figure}[ht]
\centerline{\includegraphics[width=8.1cm]{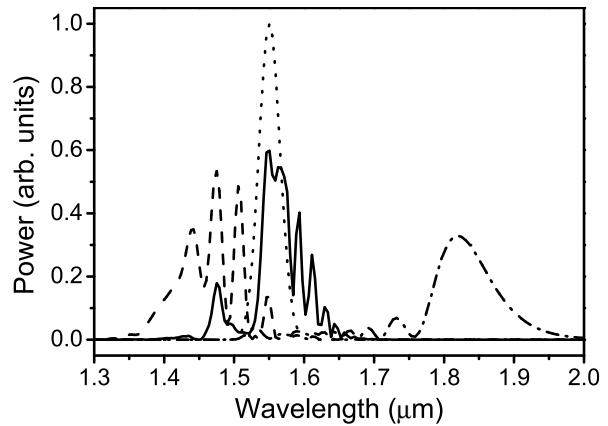}}
\caption{\label{compensation_fig} Pulse spectrum after propagation in fiber without pre-compensation (dash-dotted line) and after cascade pre-compensation stage (dashed line) and subsequent propagation through fiber (solid line). Dots indicate the launched pulse spectrum.}
\end{figure}

In high-energy short-pulse fiber amplifiers, the nonlinear phase shift can be controlled reasonably well by the technique of chirped-pulse amplification, and as a result an equally important limitation to pulse energy is stimulated Raman scattering \cite{Almantes_CPA_Review}. The red-shifts produced by Raman scattering can be compensated by blue-shifting the pulses prior to, or following, propagation in fiber. As an example, we calculate the pre-compensation of the Raman-induced red-shift of a 100-fs, transform-limited pulse centered at 1550 nm in standard single-mode fiber (modal area of $80\;{\rm \mu m^2}$ and GVD of $-23\;{\rm ps^2/km}$). The pulse energy is 1 nJ. The quadratic medium is a 4 cm-long wave-guide written in PPLN. The modal area of the waveguide is $40\;{\rm \mu m^2}$ and the GVD for the FF is $100\;{\rm ps^2/km}$ \cite{PPLN}. The phase-mismatch is set to $\Delta k = 20 \pi/\mathrm{mm}$. The pulse is first blue-shifted in the PPLN waveguide and then propagates in the fiber. These calculations indicate that the central wavelength can be kept at 1550 nm, following propagation in up to 50 cm of fiber. If no pre-compensation is utilized, the pulse is red-shifted to 1800 nm (Fig.~\ref{compensation_fig}). This result nicely complements the previously-established conclusion that the cascade nonlinearity can be used to compensate the nonlinear phase shift produced by the electronic Kerr nonlinearity under similar conditions \cite{NLM_paper}. 

Other potential applications include devices that convert peak power to frequency-shift, which can be used to switch wavelength channels or intensity discrimination with the addition of a frequency filter \cite{Xu-Liu}. 

We consider the application to pulse compression in some detail. For pulse energies in excess of 1 mJ, methods based on cubic nonlinearity for the generation of extra bandwidth cannot be used, owing to the limitations of excessive nonlinearity in single-mode waveguides and material damage through self-focusing for unguided geometries. Self-defocussing nonlinearities in quadratic media address these difficulties \cite{cascade_compression, cascade_soliton_compression}. The generalization of this approach to include frequency shifts as described here enables us to implement an analog of Raman-soliton compression \cite{Raman-soliton_compression}: high-order solitons are formed, producing a compressed primary pulse that undergoes a continuous self-frequency shift. An advantage of this approach is that the pedestal commonly produced by Raman-soliton compression consists mainly of unshifted frequency components. These components can be eliminated with a frequency filter to yield a pedestal-free pulse.

\begin{figure}[ht]
\centerline{\includegraphics[width=8.1cm]{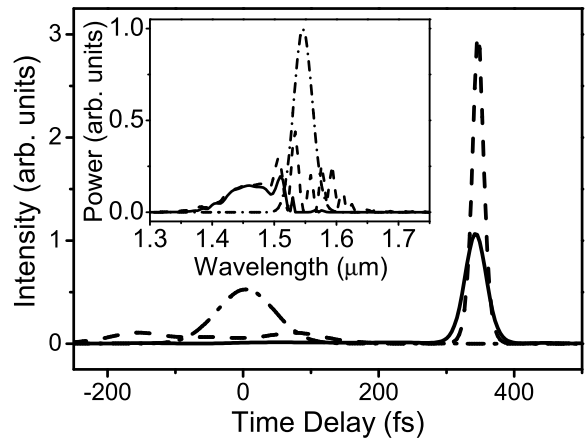}}
\caption{\label{compression_fig} Temporal profile of compressed pulses before (dashed line) and after spectral filtering (solid line) of the unshifted frequencies. Inset displays the compressed pulse spectrum before (dashed line) and after filtering (solid line). Dash-dotted lines indicate the launched temporal profile/spectrum.}
\end{figure}

Numerical simulations for realistic parameters demonstrate the utility of this approach: 100-fs, 0.6-$\rm{nJ}$ pulses with $\Delta k = 50 \pi/\mathrm{mm}$ compress to 20-fs upon propagation through a 6 cm-long waveguide in PPLN (Fig.~\ref{compression_fig}). The pulse quality ($Q_{c}$), defined as the ratio of energy contained within the FWHM of the pulse to that of the initial pulse, is calculated to be 0.65. The unshifted components can be filtered out to produce a longer (38 fs) but much cleaner ($Q_{c} = 0.91$) pulse. Compression in a second 2.5-cm-long PPLN crystal generates a 15 fs pulse with virtually no additional degradation in pulse quality ($Q_{c} = 0.90$). The resulting pulse after two stages of compression contains $\sim$50\% of the launched pulse energy. Similarly, calculations indicate that compression factors of up to 3 should be attainable with 1 mJ pulses and using a bulk BBO crystal at 800 nm, and experiments are underway to verify this compression. In addition to the high pulse quality, a practical advantage of this approach is that larger nonlinear phase shifts can be produced at the smaller phase mismatches needed in comparison to compression in quadratic media under nearly-stationary conditions. 

\section{Conclusion}

In summary we have demonstrated a new capability of cascaded quadratic processes under phase-mismatched conditions: Effectively retarded or advanced nonlinear phase shifts can be impressed on a pulse in the presence of significant GVM between the FF and SH frequencies. The frequency-domain manifestation of this non-instantaneous nonlinear response is red- or blue-shifts of the pulse spectrum. The direction and the magnitude of the frequency-shift is controllable by the choice of the phase-mismatch. Just as effectively instantaneous phase shifts from cascaded processes are analogous to bounded electronic ($\chi^{(3)}$) nonlinearities for negligible GVM, these non-instantaneous phase shifts in the presence of strong GVM are analogous to nuclear (Raman-induced) nonlinearities.

We expect the unique features of these processes to find many applications. Here, we numerically demonstrated compensation of Raman-induced frequency-shifts and high-quality pulse compression assuming typical conditions for femtosecond pulses in common quadratic nonlinear media. More generally, however, spectral shifts from cascaded quadratic processes should be applicable to all processes involving Raman-induced frequency-shifts, but with the added freedom of sign and magnitude control. 

\section{Acknowledgements}

This work was supported by the National Science Foundation under grants PHY-0099564 and ECS-0217958, and by the National Institutes of Health under grant EB002019. We acknowledge valuable discussions with B. A. Malomed and L. Torner, and we thank E. Kavousanaki for help with the numerical simulations.

\end{document}